\documentclass[apj]{emulateapj}

\usepackage{url}

\begin{document}

%\title{A Time-dependent MHD Simulation of Star-planet Interaction in the HD 189733 Planetary System}
\title{Magnetospheric Structure and Atmospheric Joule Heating of Habitable Planets Orbiting M-dwarf Stars}

\author{O. Cohen\altaffilmark{1}, J.J. Drake\altaffilmark{1}, A. Glocer\altaffilmark{2}, C. Garraffo\altaffilmark{1},
K. Poppenhaeger\altaffilmark{1}, J.M. Bell\altaffilmark{3}, A.J. Ridley\altaffilmark{4}, and T.I. Gombosi\altaffilmark{4}}

\altaffiltext{1}{Harvard-Smithsonian Center for Astrophysics, 60 Garden St. Cambridge, MA 02138, USA}
\altaffiltext{2}{NASA/GSFC, Code 673 Greenbelt, MD 20771, USA.}
\altaffiltext{3}{Center for Planetary Atmospheres and Flight Sciences, National Institute of Aerospace, Hampton, VA 23666, USA}
\altaffiltext{4}{Center for Space Environment Modeling, University of Michigan, 2455 Hayward St., Ann Arbor, MI 48109, USA}

\begin{abstract} 

We study the magnetospheric structure and the ionospheric Joule Heating of planets orbiting M-dwarf stars in the habitable zone using a set of magnetohydrodynamic (MHD) models. The stellar wind solution is used to drive a model for the planetary magnetosphere, which is coupled with a model for the planetary ionosphere. Our simulations reveal that the space environment around close-in habitable planets is extreme, and the stellar wind plasma conditions change from sub- to super-Alfv\'enic along the planetary orbit. As a result, the magnetospheric structure changes dramatically with a bow shock forming in the super-Alfv\'enic sectors, while no bow shock forms in the sub-Alfv\'enic sectors. The planets reside most of the time in the sub-Alfv\'enic sectors with poor atmospheric protection. A significant amount of Joule Heating is provided at the top of the atmosphere as a result of the planetary interaction with the stellar wind. For the steady-state solution, the heating is about 0.1-3\% of the total incoming stellar irradiation, and it is enhanced by 50\% for the time-dependent case. The significant Joule Heating obtained here should be considered in models for the atmospheres of habitable planets in terms of the thickness of the atmosphere, the top-side temperature and density, the boundary conditions for the atmospheric pressure, and particle radiation and transport.
  
\end{abstract}


%%%%%%%%%%%%%%%%%%%%%%%%%%%%%%%%%%%%%%%%%%%%%%%%%%%%%%%%%%%%%%%%%%%%%%%%%%%%%%
% Section - Introduction
%%%%%%%%%%%%%%%%%%%%%%%%%%%%%%%%%%%%%%%%%%%%%%%%%%%%%%%%%%%%%%%%%%%%%%%%%%%%%%

\section{INTRODUCTION}
\label{sec:Intro}

The simple definition of planet habitability (the ability of a planet to sustain life) is whether the surface temperature of the planet allows water to exist in a liquid form \citep{Kasting93}. The corresponding "Habitable Zone" (HZ hereafter) is the range of possible distances from the star at which a planet can have liquid surface water. This range depends primarily on the luminosity of the host star, but it can also depend on atmospheric and planetary processes that can affect the planetary surface temperature \citep[e.g.,][]{Tian05,Cowan11,Heller11,vanSummeren11,Wordsworth13}. While this intuitive definition of habitability is based on our familiarity with common life on Earth, there is growing evidence that life can arise in places and in forms we do not expect. Examples of such life forms or "Extremophiles" have been found on Earth under very cold and hot temperatures, very high pressure, high salinity, high and low pH levels, high radiation levels, and in oxygen-poor environments \citep[e.g., see][]{Rothschild01}. 

The above definition of habitability means that the search for habitable planets is focused on Earth-like, rocky planets inside the HZ. These planets are most likely to be found around M-dwarf stars, which have low luminosity so that the HZ is very close to the star, and close enough so that planets can be detected with current observational techniques. Recent surveys using the {\it Kepler} database have identified potential Earth-like planets in the HZ , taking into account the stellar luminosity, as well as atmospheric effects such as green house gasses and cloud coverage \citep[e.g.,][]{Dressing13,Gaidos13,Kopparapu13a,Kopparapu13b,Petigura13,Zsom13}.

M-dwarf stars may be the most feasible targets for detecting planets in the HZ. However, these stars are typically 
highly active magnetically, and as a fraction of their bolometric luminosity they emit more strongly at UV, EUV and X-ray wavelengths than stars of earlier spectra types \citep{Preibisch05}. If the planets are located very close to the star (as the HZ definition requires), these close-in planets can suffer from atmospheric evaporation due to the extreme EUV and X-ray radiation \citep[e.g.][]{Lammer03,Baraffe04, Baraffe06,Tian05,Garcia-Munoz07,Penz08,Yelle08,murray-clay09}, as well as from atmospheric stripping by the extreme stellar wind and Coronal Mass Ejections (CMEs) \citep{Khodachenko07,Lammer07}. In order to sustain its atmosphere, a close-in planet must have a strong internal pressure that opposes the stripping. Such a pressure can be provided by either a very thick atmosphere, similar to that of Venus, or a strong intrinsic magnetic field such as that of the Earth.  

The dynamics and energetics of planetary upper atmospheres are dominated by the interaction of the planetary magnetic field and magnetosphere with the stellar wind, in the case of a strong planetary magnetic field, or by the direct interaction of the atmosphere with the stellar wind, in the case of a weak field. The pressure balance between the planetary atmosphere and the wind depends on the dynamic and magnetic pressure of the wind, and on the atmospheric thermal and magnetic pressure. In addition, the orientation of the magnetic field of the wind compared to that of the planetary field dictates the energy transfer from the wind to the planet, as it drives magnetic reconnection which leads to particle acceleration and particle precipitation at the top of the atmosphere \citep[e.g.,][]{RussellKivelson95,Gombosi99}. Fields that drive the ions against the neutrals result in Joule Heating.  The particle precipitation can impact the local ionization and alter Joule Heating processes  \citep[see e.g.,][with references therein]{Roemer69,Hays73,Deng11} and atmospheric line excitation \citep[i.e., auroral excitation][]{Chamberlain61,Akasofu81,RussellKivelson95,Gombosi99,Paschmann02,SchunkNagy04}. We emphasize that this ionospheric Ohmic dissipation is different to the Ohmic dissipation used to explain the inflation of hot jupiters \citep[see e.g.,][]{Batygin10}. The latter occurs deeper in the atmosphere and is driven by the planetary magnetic field and the strong zonal winds observed in hot jupiters \citep{Showman08,Showman09} 

\cite{Cohen11a} investigated the plasma environment and the star-planet magnetic interaction using a global magnetohydrodynamic (MHD) model for the stellar corona and the stellar wind. In their simulation, the planet was imposed as an additional boundary condition that mimics the planetary density, temperature, and magnetic field. In a similar manner,  \cite{Cohen11b} studied the impact of a CME on the atmosphere of a close-in planet. However, in these simulations, the detailed magnetospheric structure and the energy input in to the upper atmosphere as a result of the direct interaction between the planet and the stellar wind could not be investigated.  

In this paper, we present a detailed study of the magnetospheric structure and the energy deposition into the upper atmosphere in close-in Earth-like planets orbiting an M-dwarf star. We use the upstream stellar wind conditions, extracted along the planetary orbit from a model for the stellar wind, to drive an MHD model for the global planetary magnetosphere and the ionosphere. We study how the dynamics and energetics of the planetary magnetosphere and ionosphere changes as a function of the stellar wind parameters, dynamic pressure, magnetic field topology, planetary field strength, and ionospheric conductance. In addition, we investigate how the transition along the planetary orbit between sub- to super-Alfv\'enic regime affects the magnetosphere and the energy deposition onto the planet. The Alfv\'enic point is defined by the Alfv\'enic Mach number, $M_A=u_{sw}/v_A$, which is the ratio between the stellar wind speed, $u_{sw}$, to the local Alfv\'en speed, $v_A=B/\sqrt{4\pi\rho}$, with $B$ being the local magnetic field strength and $\rho$ being the local mass density. Such a transition is unique for close-in exoplanets and does not exist in the solar system, where all the planets are almost always located in a super-Alfv\'enic solar wind flow. 

In Section~\ref{sec:Data}, we describe the particular systems we study and in Section~\ref{sec:Model} we describe our numerical approach. We describe the results in Section~\ref{sec:Results} and discuss their implications in Section~\ref{sec:Discussion}. We conclude our findings in Section~\ref{sec:Conclusions}.

%%%%%%%%%%%%%%%%%%%%%%%%%%%%%%%%%%%%%%%%%%%%%%%%%%%%%%%%%%%%%%%%%%%%%%%%%%%%%%
% Section - Data
%%%%%%%%%%%%%%%%%%%%%%%%%%%%%%%%%%%%%%%%%%%%%%%%%%%%%%%%%%%%%%%%%%%%%%%%%%%%%%

\section{SELECTED PLANETARY AND STELLAR SYSTEMS}
\label{sec:Data}

In principle, the study presented in this paper is rather generic and examines the fundamental response of Earth-like planets orbiting M-dwarf stars to the energy input from the stellar wind. Recently, \cite{Dressing13} identified three candidate Earth-like planets inside the HZ of M-dwarf stars: 1) Kepler Object of Interest (KOI) 2626.01; 2) KOI 1422.02; and 3) KOI 854.01. We choose to use the known parameters of these planet candidates (shown in Table~\ref{table:t1}) to represent three typical Earth-like planets orbiting an M-dwarf star. The magnetic fields of these planets are unknown so we assume an Earth-like magnetic field of 0.3G for all planets (with the exception of modifying the field of Planet A as described in Section 4.3). From this point, we refer
to our planet cases as ``Planet A" (using the parameters of KOI 2626.01), ``Planet B" (with the parameters of KOI 1422.02), and ``Planet C" (with the parameters of KOI 854.01), where the planets are ordered according to their  distance from the star (Planet A being the closest).

The stellar wind model (described in Section~\ref{SCsolution}) is driven by data describing the photospheric radial magnetic field (magnetograms). Such data are not available for any of the above systems. However, several observations of stars with similar parameters to those we are interested in have been made using the Zeeman-Doppler imaging (ZDI) method \citep{DonatiSemel90}. These observations enable construction of surface maps of the large-scale stellar magnetic field, which can be used to drive our
model for the stellar wind. It is important to mention that the validity of ZDI data has been questioned, since this reconstruction process does not take into account the Stokes components that cannot be measured, and they do not account for the small-scale magnetic field that may be significant \citep{Reiners09}.  \citet{Garraffo13} have shown that missing small-scale flux can have a significant effect on the predicted X-ray emission, but that the wind solutions, that are the primary interested here, are much less sensitive to this and instead depend more strongly on the large-scale field.

\cite{Morin08} have constructed ZDI maps for a number of mid M-dwarf stars. We have identified the star EV Lac, a mid-age M3.5 class star, as a star with the most similar parameters to the systems we are interested here (in particular, the effective temperature, $T_{eff}$, while the rotation periods of these systems are currently unknown).  We have constructed a magnetic map based on that of  \cite{Morin08}, assuming it does not contain much small-scale structure and it represents essentially a tilted dipole with a polar field strength of 1.5-2~kG.

We stress that we choose this approach in order to generate a solution with azimuthally varying plasma conditions along the planetary orbit based on typical parameters of M-dwarf stars. This could not be achieved by using an aligned dipole, which would yield a symmetric and constant solution. Here we assume that the stellar wind parameters for EV Lac are similar to those of our planets, and that these parameters can be used to study the effects of the stellar wind on the planetary magnetosphere and ionosphere. We are interested in the tentative effects on the planet due to the {\it close proximity of the planet to the star}, the high dynamic pressure, and the strong field of the stellar wind.  The use of the parameters of an active M dwarf such as EV Lac (instead of the unknown parameters or idealized parameters) represents a reasonable and tractable approach.

%%%%%%%%%%%%%%%%%%%%%%%%%%%%%%%%%%%%%%%%%%%%%%%%%%%%%%%%%%%%%%%%%%%%%%%%%%%%%%
% Section - Model
%%%%%%%%%%%%%%%%%%%%%%%%%%%%%%%%%%%%%%%%%%%%%%%%%%%%%%%%%%%%%%%%%%%%%%%%%%%%%%

\section{NUMERICAL SIMULATIONS}
\label{sec:Model}

For our simulations, we use the generic {\it BATS-R-US} magnetohydrodynamic (MHD) code \citep{powell99} and the Space Weather Modeling Framework \citep[SWMF,][]{toth05,Toth12}, which were developed at the Center for Space Environment Modeling at the University of Michigan. The SWMF provides a set of models for different domains in space physics, such as the solar corona, the inner and outer heliosphere (i.e., the interplanetary environment and the solar wind), planetary magnetosphere, planetary ionospheres, and planetary upper atmospheres. These codes are generally speaking based on solving the extended MHD or electrodynamic equations. In addition, the SWMF includes codes for the planetary radiation environment, which are particle codes. All of these models (or part of them) can be coupled together to provide solutions for the space environment that are much more detailed and physics-based than any solution provided by each of these models independently. Our modeling approach is based on a large number of studies of planets in the solar system carried out with the SWMF in a similar manner (see publications list at the \url{http://csem.engin.umich.edu}).

In this work, we use the Stellar Corona (SC) MHD code to obtain the solution for the interplanetary environment of the three planet candidates. We then use these solutions to drive a Global Magnetosphere (GM) MHD model for these planets. In order to obtain a more realistic magnetospheric solution, as well as calculating the energy input at the top of the upper atmosphere, we couple the GM model with a model for the planetary Ionospheric Electrodynamics (IE). In the next sections, we describe in detail each model and the coupling procedure. 

\subsection{STELLAR WIND MODEL} 
\label{SCsolution}

In order to obtain a solution for the stellar wind, we use the SC version of {\it BATS-R-US} \citep{Oran13,Sokolov13,Vanderholst14}. The model is driven by the photospheric radial magnetic field (see Section~\ref{sec:Data}), which is used to calculate the three-dimensional potential magnetic field above the stellar surface \citep{altschulernewkirk69}.  The potential field solution is in turn used as the initial condition for the magnetic field in the simulation domain. Once the initial potential field is determined, the model calculates self-consistently the coronal heating and the stellar wind acceleration due to Alfv\'en wave turbulence dissipation, taking into account radiative cooling and the electron heat conduction. Unlike most MHD models for the solar corona, the lower boundary of this model is set at the chromosphere, so that it does not initially assume a hot corona at its base.  Instead, the heating is calculated self-consistently. 

While the model works very well for the Sun, applying this model to M-dwarf stars is not so trivial.  Our reference star, EV~Lac, has stronger magnetic fields than the Sun, and there is a general lack of observations of the winds of other stars \citep[see, e.g.,][and references therein]{wood04,gudel07}. Our basic assumption is that the winds of M-dwarfs are accelerated in a similar manner to the Sun, with a combination of thermal acceleration (taken into account in the model) and dissipation of magnetic energy.  An example of a different mechanism that the model cannot account for is stellar winds from highly evolved giants, which are likely driven by radiation pressure on dust grains \citep[e.g.][]{lamerscassinelli99} or by radial pulsations \citep[e.g.][]{Willson00}.  Based on our assumption, several studies of Sun-like stars have been made using the SC model. \cite{cohen10b} have applied the model to the active star AB Doradus, where they argued for the validity of the model for solar analogs. Similarly, \cite{Cohen11a} performed simulation of HD 189733 driven by ZDI observations reproduced from \cite{fares10}. The model has been used recently by \cite{Cohen14} to perform a parametric study on the stellar wind dependence on magnetic field strength, base density, and rotation period. 

As stated above, we use the ZDI observation of EV Lac, as well as its stellar parameters of $R_\star=0.3R_\odot$, $M_\star=0.35M_\odot$, and rotation period, $P_\star=4.3$ days \citep{Morin08}, to drive the SC model.  We also adopt the  bolometric luminosity $L_{bol}=4.5\cdot10^{24}\;W$ from \citet{Morin08}, which corresponds to an effective temperature of $3400\;K$.  We use a spherical grid that extends up to $100R_\star$ so as to include the orbits of all three planets. Once a steady-state solution is obtained, we extract the stellar wind parameters of number density, $n$, velocity, $\mathbf{u}$, magnetic field, $\mathbf{B}$, and plasma temperature, $T$, at a given point along the orbit of one of the planets. 

The SC solution is provided in the frame of reference rotating with the star (HelioGraphic Rotating coordinates or HGR). In this coordinate system, the $\hat{Z}$ axis is aligned with the rotation axis of the star, the $\hat{X}$ axis is aligned with the initial time of the ZDI observation (longitude "0"), and the $\hat{Y}$ axis completes the right-hand system. The GM model uses the Geocentric Solar Magnetospheric (GSM) coordinate system, which is identical to the Geocentric Solar Ecliptic (GSE) system for the case of a planetary dipole perpendicular to the ecliptic plane. 
%This is our assumption for all simulations presented here, where we only account for the strength of the planetary dipole magnetic field as a free parameter. 
%In this case, 
In this special case, the planetary GSE/GSM coordinate system is defined with $\hat{X}$ pointing from the planet to the star (negative radial direction in the stellar frame of reference assuming a circular planetary orbit), $ \hat{Z}$ is pointing to the north pole of the planet (perpendicular to the ecliptic plane and the plane of orbit), and $\hat{Y}$ completes the right-hand system. With a circular planetary orbit and an aligned planetary dipole, the conversion between the coordinate systems is:

\begin{eqnarray}
&X_{GSE}=-r_{HGR} &\\
&Y_{GSE}=\phi_{HGR} &\nonumber\\
&Z_{GSE}=\theta_{HGR}.& \nonumber
\end{eqnarray}
The orbital speed of the planet, $U_{orb}$, could be easily considered as a constant addition to $U^{GSE}_y$. However, it is hard to estimate what would be the change in $B^{GSE}_y$, which has a strong effect on the magnetospheric current system. After carefully confirming that the Alfv\'en Mach number with and without the addition of $U_{orb}$ is essentially the same for all cases, we have decided to exclude this motion from our simulation. 

\subsection{GLOBAL MAGNETOSPHERE AND IONOSPHERE MODELS }
\label{sec:GMIE}

The Global Magnetosphere (GM) model solves the MHD equations on a Cartesian grid for the physical domain that includes the planet as the inner boundary, and the extent of the planetary magnetosphere. The model is driven from the outer boundary that is facing the star by the upstream stellar wind conditions, which can be fixed or time-dependent. This boundary is defined with inflow boundary conditions for all the MHD parameters for the case of super-Alfv\'enic stellar wind conditions. In the case of sub-Alfv\'enic stellar wind conditions, the boundary conditions for the pressure are changed to float in order to diminish numerical effects on the boundary from the inner domain (this does not happen for super-Alfv\'enic boundary conditions). For the same reason, we also set the upstream boundary very far from the planet. The boundary conditions for all the other  boundaries are set to float for all the MHD parameters. 

The inner boundary in GM is defined by the planetary parameters of radius, mass, magnetic field, and density \cite[as described in][]{powell99}. In order to better constrain the velocities at the inner boundary, the GM model is coupled with a model for the Ionospheric Electrodynamics (IE), which is a completely separate model that solves for the electric potential in the ionosphere. The IE model provides the convection electric field, which is then used to calculate the velocities at the inner boundary of GM, along with the rotational velocity of the planet.

The coupling procedure, which is described in detail in \cite{Ridley04} and in \cite{toth05}, begins by calculating the field-aligned currents, $\mathbf{J}_\parallel=(\nabla\times\mathbf{B})\cdot\mathbf{b}$, at $3R_p$ in the GM model. The currents are then mapped assuming a dipole planetary field to the ionospheric height of 120 km, using a scaling of $B_I/B_3$, where $B_I$ and $B_3$ are the field strengths at the ionosphere and at the point of origin at $3R_p$.
Using the electric conductance tensor, $\Sigma$, an electric potential is solved with:
\begin{equation}
J_r(r_I)=\left[  \nabla_\perp \cdot (\Sigma\cdot \nabla\Psi) _\perp  \right]_{r=r_I},
\end{equation}
and this potential is mapped to the inner boundary of the GM at $2R_p$ (for numerical efficiency, the inner boundary is set higher than $1R_p$ to reduce the planetary dipole strength and increase the numerical time step). At the final stage, the electric field, $\mathbf{E}=-\nabla\Psi$, and the bulk convection velocity, $\mathbf{V}=\mathbf{E}\times\mathbf{B}/B^2$ are calculated. The velocity field (along with the rotational velocity) is applied at the GM inner boundary.

The coupling of GM and IE models enables us to to specify more realistic inner boundary conditions for the GM.  This improved boundary specification allows us to estimate the energy input at the top of the planetary atmosphere due to its interaction with the stellar wind and the precipitating particles. Assuming the currents and the (scalar) conductance, $\sigma$, are known in the IE model, it is trivial to calculate the Joule Heating:
\begin{equation}
Q=\mathbf{J}\cdot \mathbf{E}=J^2/\sigma, 
\end{equation}
from the generalized Ohm's law with $\mathbf{J}=\sigma\mathbf{E}$. 

We note that, in a way, the electric conductance captures all the atmospheric parameters in it (chemistry, photoionization etc.). For the case of the Earth, a more complex conductance can be used. The conductance depends significantly on the solar EUV flux  \citep[and correlates with the radio flux in the F10.7 centimeter wavelength, see e.g.,][]{Moen93}, auroral particle precipitation \citep[e.g.,][]{Robinson87,Fuller-Rowell87}, and other processes that are quite dependent on understanding the environment near the planet. Global models of the upper atmosphere, such as the one described by \cite{Ridley02} can also be used, but they are driven by the observed solar luminosity and auroral precipitation, so it is difficult to drive them properly for other planets to determine the conductance patterns. In order to avoid further uncertainties, here we choose to use a constant Pedersen conductance, $\sigma_p$, of the order of the one used for the Earth \citep[e.g.,][]{Ridley04}. The Pedersen conductance allows the magnetospheric currents to close through the ionosphere, and it depends on the collision frequency between electrons and ions, $\nu_{e,i}$,  and the electron plasma frequency, $\Omega^2_e=n_e e^2/\varepsilon_0 m_e$:
\begin{equation}
\sigma_p=\frac{\nu^2_{e,i}}{\nu^2_{e,i}+\Omega^2_e} \sigma_0,
\end{equation}
with $\sigma_0=n_e e^2/\nu_{e,i}m_e$, where $n_e$ is the electron density, $m_e$ is the electron mass, $e$ is the electric charge, and $\varepsilon_0$ is the permittivity in free space. The extreme EUV irradiation of close-in planets around M-dwarf stars should reduce the altitude of their ionospheres to regions with higher electron density. It is not trivial to predict how the Pedersen conductance will change as a result of the increased EUV radiation, as it has a complicated dependence on the density variations of ions electrons and neutrals, ionization rates, atmospheric chemistry, and perhaps other factors. In the results section, we probe the sensitivity of our calculations to this by showing how a simple increase in the Pedersen conductance affects the Joule Heating for a given set of parameters. 

%%%%%%%%%%%%%%%%%%%%%%%%%%%%%%%%%%%%%%%%%%%%%%%%%%%%%%%%%%%%%%%%%%%%%%%%%%%%%%
% Section - Results
%%%%%%%%%%%%%%%%%%%%%%%%%%%%%%%%%%%%%%%%%%%%%%%%%%%%%%%%%%%%%%%%%%%%%%%%%%%%%%

\section{RESULTS}
\label{sec:Results}

\subsection{STELLAR WIND AND CORONAL STRUCTURE}
\label{SCSol}

The model wind solution for EV~Lac shows an average speed of about 300~km~s$^{-1}$ and total mass loss rate of is about $3\times10^{-14} M_\odot$~yr$^{-1}$.  While these values are close to those of the solar wind, the mass loss rate per unit stellar surface area from this diminutive M dwarf is an order of magnitude higher.

\begin{figure*}[h!]
\centering
\includegraphics[width=7.in]{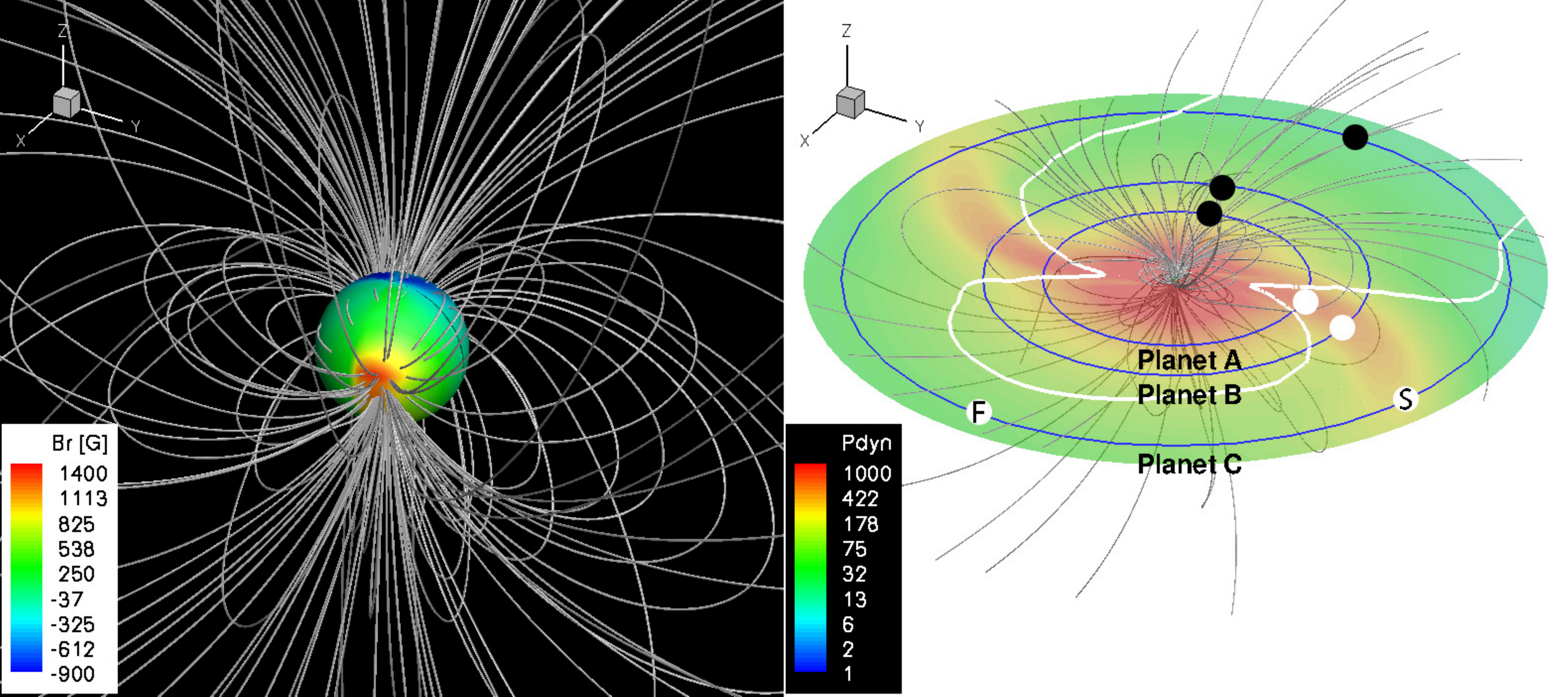}
\caption{The coronal solution for EV Lac. Left: color contours of the photospheric radial magnetic field used in the simulation (based on the ZDI map from \cite{Morin08}) is shown on a sphere representing $r=1R_\star$. Selected coronal magnetic field lines are shown in grey. Right: selected coronal magnetic field lines are shown in grey with the equatorial plain colored with contours of the ratio between the stellar wind dynamic pressure and the dynamic pressure of the ambient solar wind base on the parameters from Table~\ref{table:t3}. Also shown are the circular orbits of the three planets, and the Alfv\'en surface crossing of the equatorial plain (represented by the solid white line). The locations where the upstream conditions were extracted are marked in black circles for the sub-Alfv\'enic regions, and in white circles for the super-Alfv\'enic regions. The letters 'F' and 'S' shows the location of the fast and slow super-Alfv\'enic conditions for planet C.}
\label{fig:f1}
\end{figure*}

Figure~\ref{fig:f1} shows the steady-sate coronal and stellar wind solution for EV Lac. It shows the orbits of the three planets, selected coronal magnetic field lines, and color contours of the ratio between the dynamic pressure in the solution to that of a typical solar wind conditions at 1~AU (see background solar wind conditions in Table~\ref{table:t3}).  The dynamic pressure of the ambient stellar wind at these close-in orbits is 10 to 1000 times larger than that near Earth. In addition, the magnetic field strength ranges between 500--2000~nT along the orbit of Planet A, between 200--800~nT along the orbit of Planet B, and between 100--200 nT along the orbit of Planet C. This is in contrast to a field strength of the order of 1--10~nT for typical solar wind conditions at 1~AU. Finally, the temperature of the ambient stellar wind along the orbits of the planets ranges between 300,000 K to over 2 MK. The typical solar wind temperature is about $10^4$K.

As seen in Figure~\ref{fig:f1}, the stellar wind conditions change from sub- to super-Alfv\'enic along the orbits of all three planets. For Planet A and Planet B, we drive the GM simulation using both sub- and super-Alfv\'enic upstream conditions. For Planet C, we perform three GM simulations using sub-Alfv\'enic conditions, super-Alfv\'enic conditions with slow (more dense) stellar wind, and super-Alfv\'enic conditions with fast (less dense) stellar wind. Table~\ref{table:t2} summarizes the upstream conditions used to drive the GM model for the different Planets.

\subsection{STEADY STATE MAGNETOSPHERIC STRUCTURE}
\label{sec:SteadyStateGM}

\begin{figure*}[h!]
\centering
\includegraphics[width=7.in]{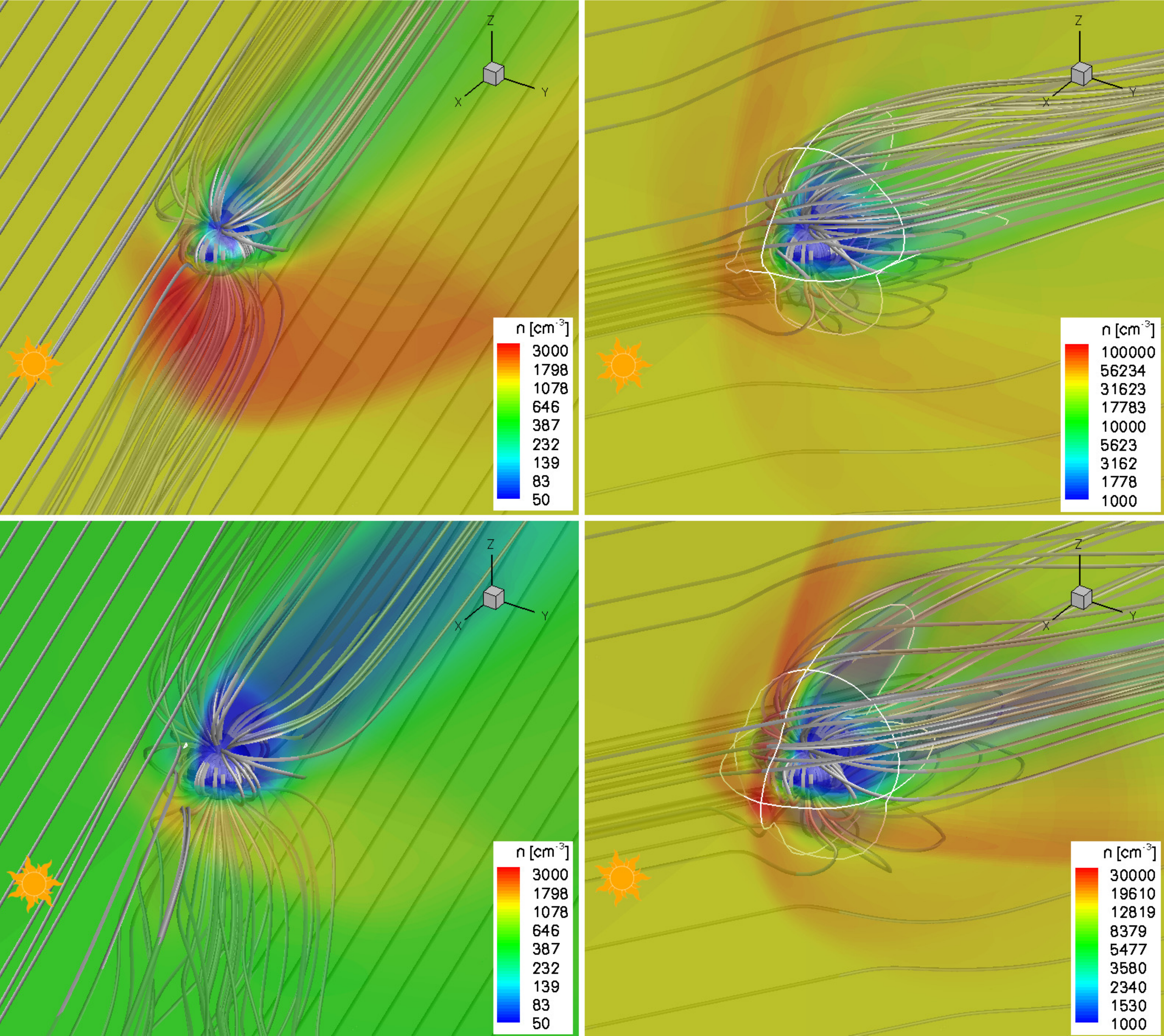}
\caption{The magnetospheres of Planet A (top) and Planet B (bottom) for sub-Alfv\'enic (left) and super-Alfv\'enic (right) stellar wind conditions. Color contours show the number density (note the different scales in the panels) and selected magnetic field lines are shown in grey. The direction of the star (the direction from which the stellar wind is coming) is marked by the small yellow Sun shape. The structures and trends are similar for Planet C.}
\label{fig:f2}
\end{figure*}

Figure~\ref{fig:f2} shows the magnetospheric structures of Planet A and Planet B (the magnetospheric structure of Planet C is qualitatively similar). The most notable result is the dramatic change in the magnetospheric topology when the stellar wind upstream conditions change from sub- to super-Alfv\'enic. For sub-Alfv\'enic conditions the planetary field lines simply merge with the stellar wind field lines in an Alfv\'en-wings topology \citep{Neubauer80,Neubauer98}. An Alfv\'en-wings configuration arises when a conducting obstacle moves in a plasma with a sub-Alfv\'enic speed and it is the result of the configuration of the current system that connects the external plasma with the currents flowing inside the body, along low flow cavities. It has been well observed and studied for the Jovian moons Io \citep{Neubauer80,Combi98,Linker98,Jacobsen07} and Ganymede \citep{Ip02,Kopp02,Jia08}. Some studies suggest that this configuration can be obtained at Earth during periods when the solar wind has a weak Alfv\'enic Mach number \citep{Ridley07,KivelsonRidley08}.

For super-Alfv\'enic upstream conditions, an Earth-like magnetospheric configuration forms with the planetary field lines being draped over by the stellar wind, and a magnetopause bow shock being forming in front of the planet along with a magnetotail behind. 

The transitioning between sub- to super-Alfv\'enic conditions occurs twice per orbit and within a short time. This has implications for the energy input into the upper atmosphere, which are discussed in Section~\ref{sec:TimeDependentKOI2626}.

\subsection{STEADY STATE JOULE HEATING}
\label{sec:SteadyStateIE}

\begin{figure*}[h!]
\centering
\includegraphics[width=5.in]{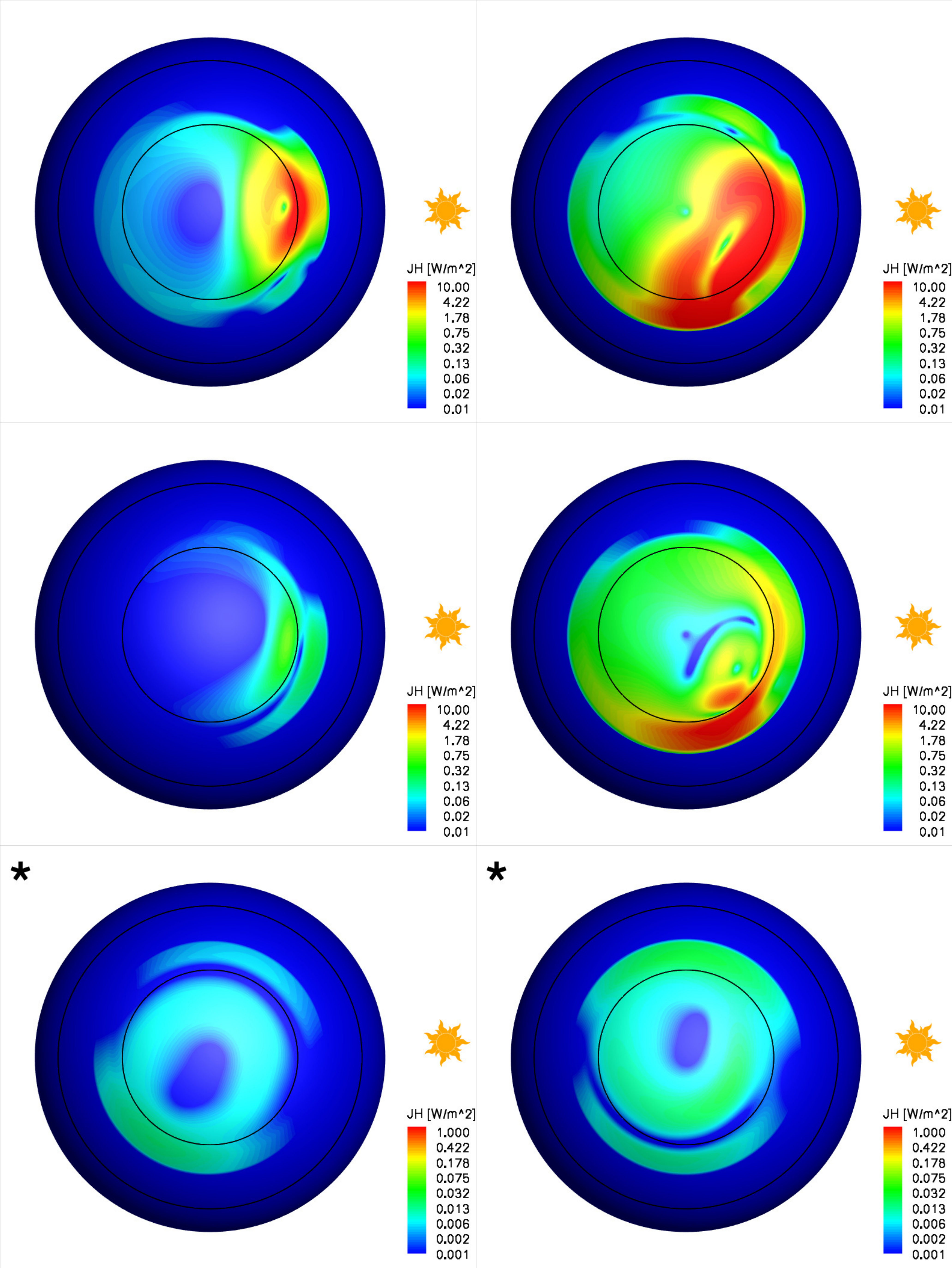}
\caption{Height-integrated Joule Heating (in $W\;m^{-2}$) of the ionospheres of Planet A (top), Planet B (middle), and Planet C (bottom) for sub-Alfv\'enic (left) and super-Alfv\'enic (right) stellar wind conditions displayed in polar plots of the planetary Northern hemisphere. The sub-stellar point (day side) is marked by the small Sun shape. Note that the heating of Planet C is displayed on a reduced color scale (marked with *). The distribution of the Joule Heating in the Southern hemisphere is similar, but is mirrored towards the night side instead of the day side.}
\label{fig:f3}
\end{figure*}

Figure~\ref{fig:f3} shows the distribution of the height integrated ionospheric Joule Heating, $Q_l$, (in units of $W\;m^{-2}$) in a format of polar plots. Each pair of panels shows the Joule Heating for one of the three planets extracted from a steady-state GM-IE simulation. The heating is clearly stronger for the closer orbit of Planet A than the more distant orbit of Planet C, for which the color scale of the plot has been extended to lower values of heating for clarity. The figure also shows that the heating is stronger for Super-Alfv\'enic stellar wind conditions than for Sub-Alfv\'enic conditions, and that the distribution is asymmetric, as expected, due to the asymmetric stellar wind magnetic field and magnetospheric field-aligned currents.

\begin{figure*}[h!]
\centering
\includegraphics[width=6.in]{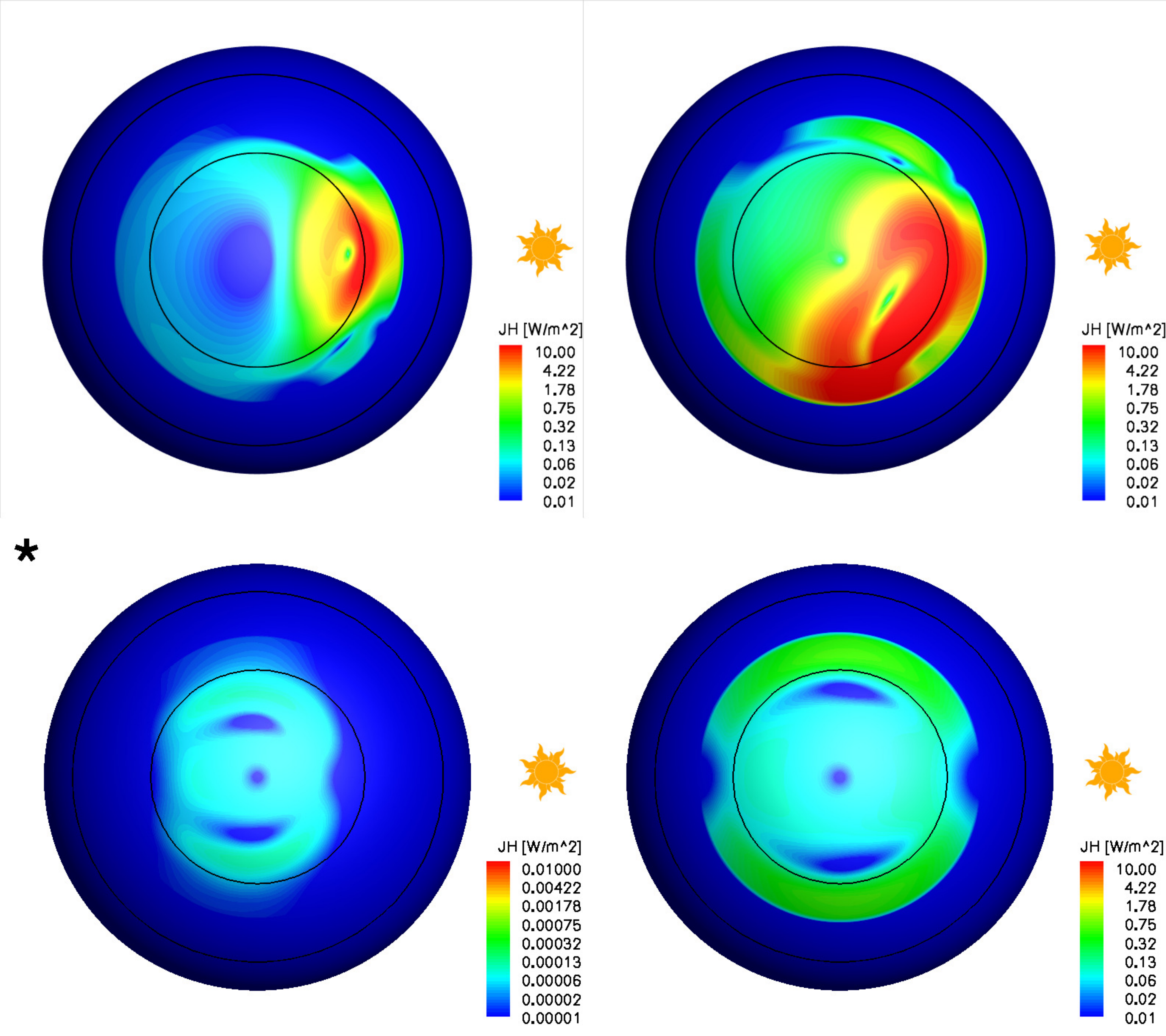}
\caption{Height-integrated Joule Heating (in $W\;m^{-2}$) of the ionospheres of Planet A (top),  and for ambient solar wind (bottom-left) and CME (bottom-right) conditions at Earth taken from Table~\ref{table:t3}. The display is similar to that of Figure~\ref{fig:f3}.}
\label{fig:f4}
\end{figure*}

For comparison with the case of the Earth, we ran the model using the planetary parameters of the Earth, and upstream conditions of a typical quiet solar wind, as well as with upstream conditions of a strong CME event. The upstream parameters of these reference runs are summarized in Table~\ref{table:t3}. Figure~\ref{fig:f4} shows the ionospheric Joule Heating for Planet A and for these reference Earth cases. It shows that the heating by the ambient stellar wind conditions at close-in orbits is about four orders of magnitude higher than the case of the ambient solar wind conditions at Earth, and is even higher than the heating during a strong space weather event on Earth. 

\begin{figure*}[h!]
\centering
\includegraphics[width=4.5in]{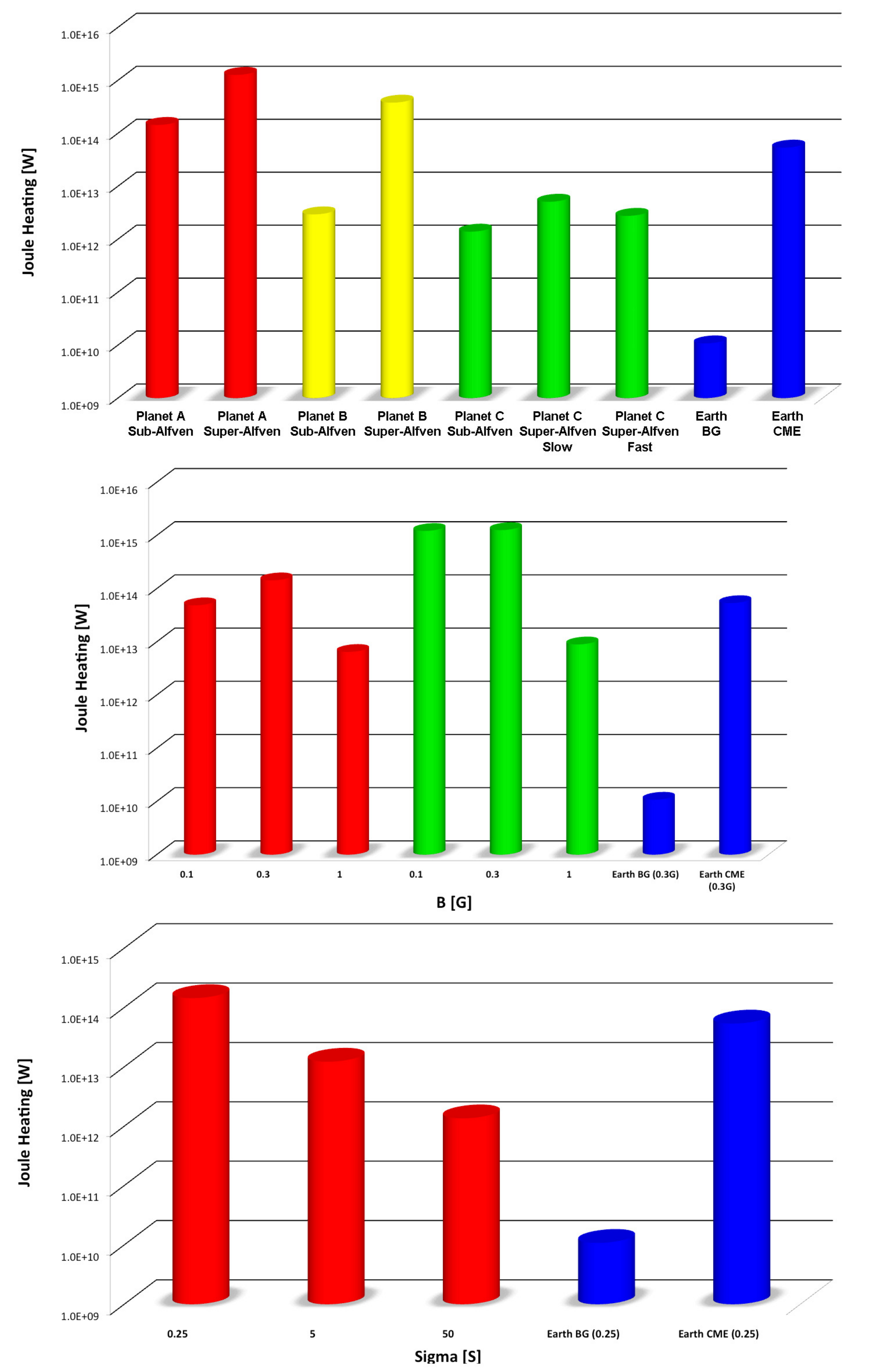}
\caption{Top: the total power (in $W$) of the integrated Joule Heating for all the planets and reference Earth cases. Middle: the total Joule Heating power for Planet A as a function of planetary magnetic field strength shown for the sub-Alfv\'enic (red) and super-Alfv\'enic (green) cases. Also shown are the reference Earth cases (blue). Bottom: the total Joule Heating power for Planet A for sub-Alfv\'enic (red) conditions as a function of the ionospheric conductance. Also shown are the reference Earth cases (blue).}
\label{fig:f5}
\end{figure*}

The top panel in Figure~\ref{fig:f5} shows the total area integrated power, $P=\int Q_l da$, for all the solutions, where $da$ is the surface element of the 2D ionospheric sphere. It is consistent with the trend which is usually seen in Figure~\ref{fig:f3}, with the power being the greatest for the closest planet. For the Planet A, the power reaches $10^{14}-10^{15}\;W$, which is 0.01-0.1\% of the total incident radiation of the M-dwarf star, assuming $L_{bol}=4.5\cdot10^{24}\;W$, $a=0.06\;AU$ ($36R_\star$), and $P\approx\frac{1}{4}L_{bol}(R_p/a)^2\approx10^{18}\;W$. Since the planetary albedo is likely not zero, much of this radiative will be reflected, and the significance of the Joule Heating can be even greater. For all cases, the heating in the super-Alfv\'enic regime is higher than the sub-Alfv\'enic one, and for the case of Planet C, the heating is greater for the super-Alfv\'enic regime with a slow stellar wind than for the super-Alfvenic upstream conditions with a fast stellar wind. This is due to the order of magnitude density variation of the stellar wind upstream conditions, which increases the wind's dynamic pressure (i.e., $\rho u^2_{sw}$). Moreover, the super-Alfv\'enic sectors are located near the helmet streamers, where the velocity component which is not parallel to the magnetic field (i.,e., the non-radial component) is greater. As a result, and the upstream electric field, $\mathbf{E}=-\mathbf{u}\times\mathbf{B}$, which dictates the ionospheric electric potential and the coupling between the stellar wind and the planetary magnetosphere, is larger as well. 

In the middle panel of Figure~\ref{fig:f5}, we show the Joule Heating power for different planetary magnetic field strengths for the sub- and super-Alfv\'enic upstream conditions in Planet A, along with the reference Earth cases. The power is greater for the weaker planetary fields due to the strong penetration by the stellar wind. As the field increases to 1G, the wind is pushed back and the energy input to the upper atmosphere is reduced. In addition, the total size of the heated polar cup is reduced too. Due to the uncertainties about the planetary magnetic field, here we assume a planetary dipole field aligned with the stellar rotation axis. We omit the effect of different dipole orientation, which could affect the energy input as a result of magnetic reconnection between the stellar wind and the planetary magnetosphere (e.g., a geomagnetic storm). 

The bottom panel of Figure~\ref{fig:f5} shows the Joule Heating power for an ionospheric conductance of 0.25 (the value used for all simulations), 5, and 50 Siemens for the sub-Alfv\'enic case of Planet A, along with the reference Earth cases. As mentioned in Section~\ref{sec:GMIE}, the actual ionospheric conductance depends on the atmospheric density, composition, the level of ionization, and the level of photoionization (the stellar EUV and X-ray flux).  The power varies inversely with the value of the conductance.  As the conductance represents the mobility of the ions, a lower value means that the ions are less mobile and collide more frequently with neutrals.  As a result, the energy dissipation (or Joule Heating) increases.

\subsection{TIME-DEPENDENT SOLUTION FOR PLANET A}
\label{sec:TimeDependentKOI2626}

The results presented in Sections~\ref{sec:SteadyStateGM} and \ref{sec:SteadyStateIE} are the steady-state solutions for the different planets, which are driven by the upstream stellar wind conditions extracted from the SC solution at particular locations along their orbit. As stated in Section~\ref{sec:SteadyStateGM}, the magnetospheric structure undergoes significant change as the planets move from the sub- to super-Alfv\'enic plasma sectors along the orbit. In order to obtain the dynamic effect of such a transition, we performed a time-dependent simulation of Planet A starting at a sub-Alfv\'enic sector and ending in a super-Alfv\'enic sector. In this time-dependent simulation, the upstream conditions in GM are updated every 2:52 hours (10 degrees of the orbit of 4.3 days) so that the change in the driving stellar wind conditions is captured (in contrast to the single set of stellar wind conditions for the steady-state).   

\begin{figure*}[h!]
\centering
\includegraphics[width=6.in, angle =0]{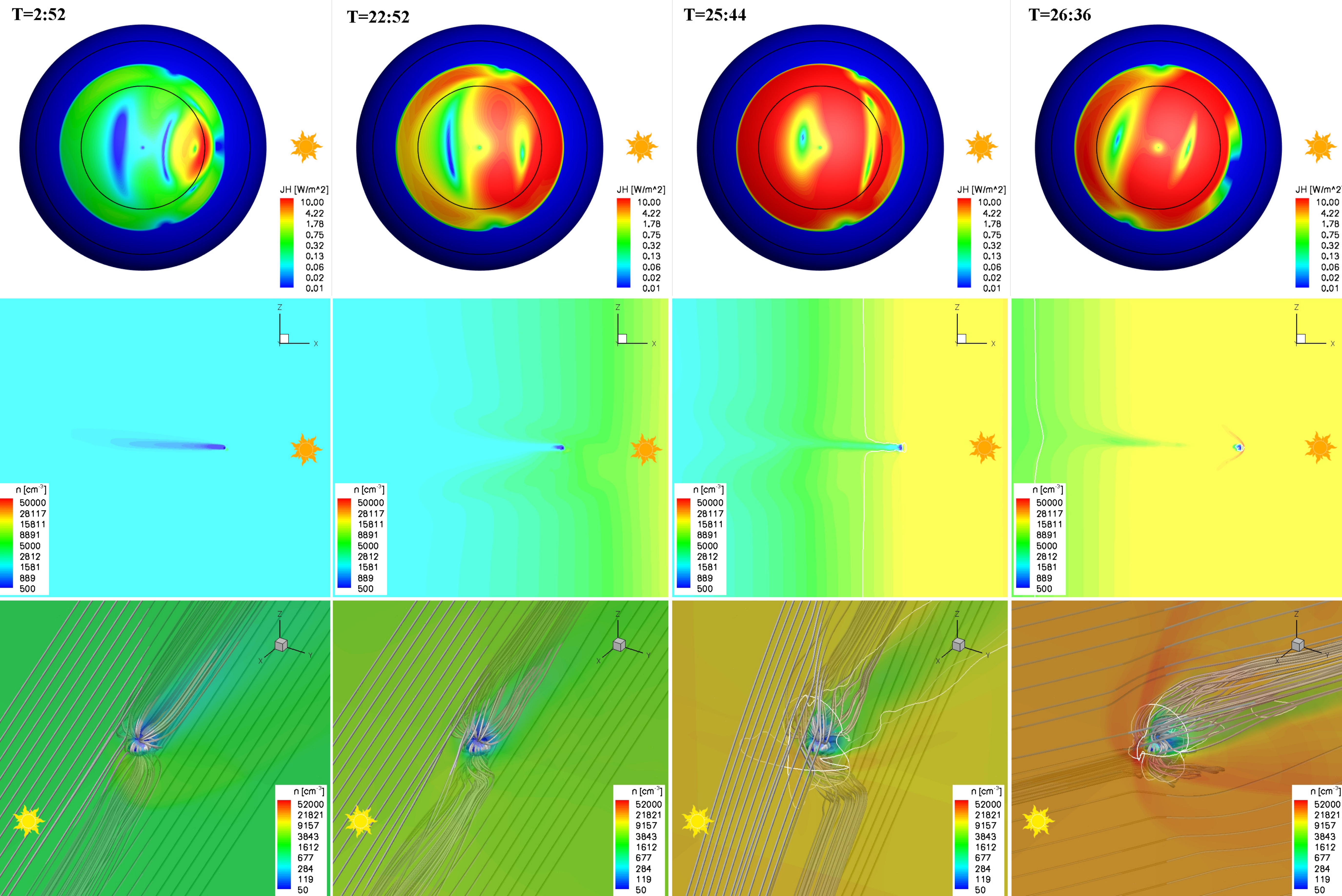}
\caption{Results from the time-dependent simulation of Planet A for 2:52h, 22:52h, 25:44h, and 26:36h. The top panels show snapshots of the ionospheric Joule Heating with a display similar to that of Figure~\ref{fig:f3}. The middle panels show y=0 cuts colored with contours of number density. The solid white line represents the $M_A=1$ line as the planet passes from the sub- to the super-Alfv\'enic sector. The bottom panel shows the time evolution of the magnetosphere with a display similar to that in Figure~\ref{fig:f2}. The direction of the star in all panels is marked by the small yellow Sun shape.}
\label{fig:f6}
\end{figure*}

Figure~\ref{fig:f6} shows the magnetospheric structure, as well as the Joule Heating in the ionosphere at after 2:52h, 22:52h, 25:44h, and 26:36h. The planet moves to the super-Alfv\'enic sector around 25:00h, with the white line in the lower two panels representing the $M_A=u_{sw}/v_a=1$ line. This line also helps to identify the magnetopause bow shock built in front of the planetary magnetosphere as the planet transitions to a super-Alfv\'enic sector. It can be clearly seen from the top four panels in Figure~\ref{fig:f6} that there is a significant heating of the upper atmosphere as the planet transitions between the sectors. 

\begin{figure*}[h!]
\centering
\includegraphics[width=6.in]{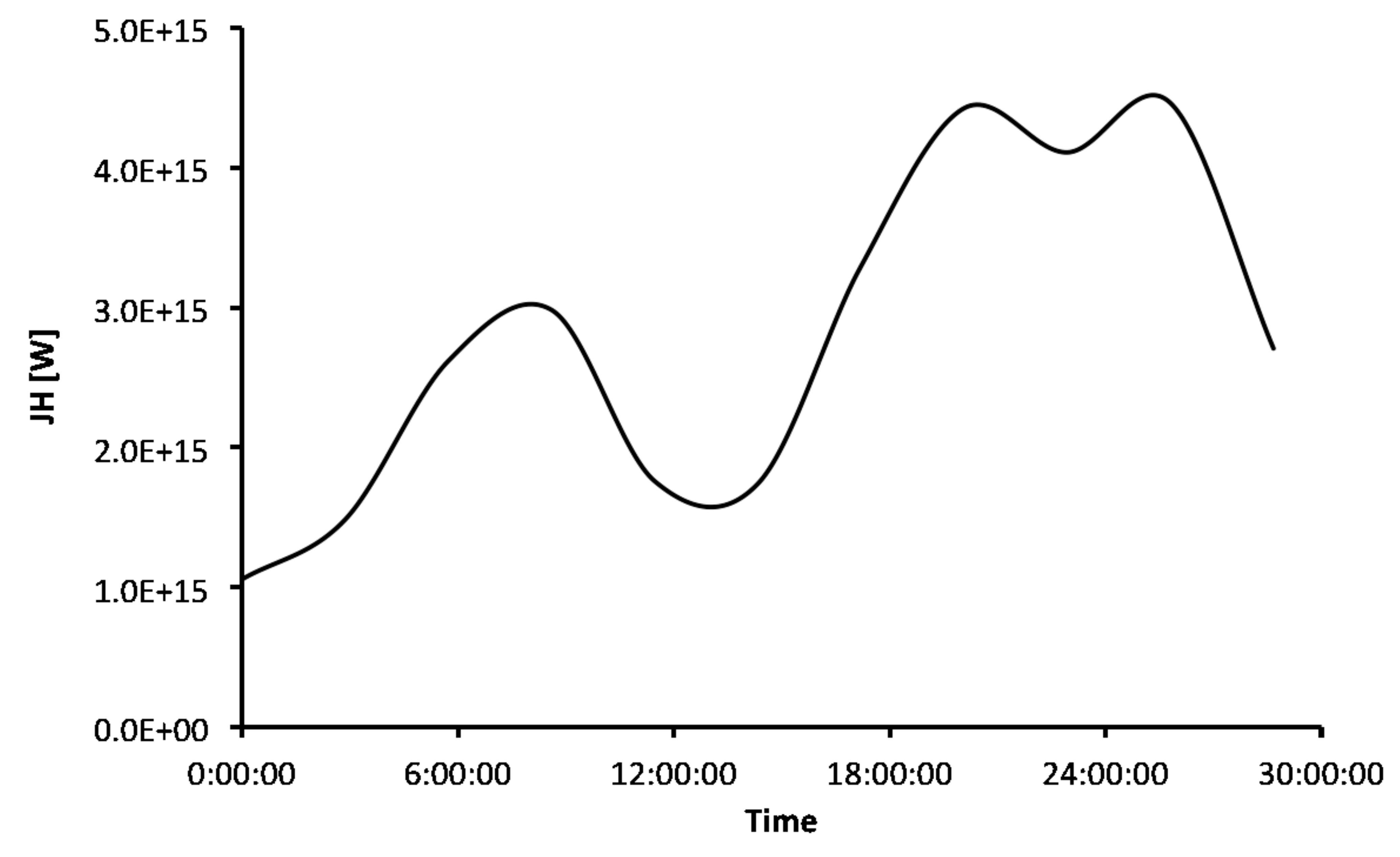}
\caption{The change in Joule Heating power vs.\ time for the time-dependent simulation of Planet A.}
\label{fig:f7}
\end{figure*}

Figure~\ref{fig:f7} shows the temporal change in the total power as a result of the planetary motion along the orbit. The total power increases by 50\% as the planet moving from the sub- to super-Alfv\'enic sector. The heating in the sub-Alfv\'enic sector is 10 times higher in the dynamic simulation compared to the steady-state obtained at the same point, and the heating in the super-Alfv\'enic sector is 50\% higher in the dynamic simulation compared to the steady-state obtained at the same point (in this case, the heating is almost 1\% of the incident stellar radiative power). This is due to the fact that the dynamic simulation captures the temporal change in the magnetic field, which drives stronger field-aligned currents (the time derivative of the magnetic field is zero for the steady state, but can and does vary in the time-dependent case).

%%%%%%%%%%%%%%%%%%%%%%%%%%%%%%%%%%%%%%%%%%%%%%%%%%%%%%%%%%%%%%%%%%%%%%%%%%%%%%
% Section - Discussion
%%%%%%%%%%%%%%%%%%%%%%%%%%%%%%%%%%%%%%%%%%%%%%%%%%%%%%%%%%%%%%%%%%%%%%%%%%%%%%

\section{DISCUSSION}
\label{sec:Discussion}

The results of our simulations reveal a number of interesting findings regarding close-in planets orbiting M-dwarf stars. These findings relate to the extreme space environment surrounding these planets and planetary shielding and protection, the change in the magnetospheric structure as a result of the planetary orbital motion, and the Joule Heating of the upper atmospheres as a result of the interaction with the stellar winds. Below we discuss in details each one of these findings.

\subsection{EXTREME SPACE ENVIRONMENT AND PLANETARY PROTECTION}
\label{sec:PlanetaryProtection}

Our results show in detail that the space environment of close-in exoplanets is much more extreme in terms of the stellar wind dynamic pressure, magnetic field, and temperature. Each of these parameters is about 1--3 orders of magnitude higher than the typical solar wind conditions near the Earth. The ultimate consequence of such an extreme environment, in the context of this paper, is the potential for stripping of the planetary atmosphere \citep[in addition to atmospheric evaporation by the enhanced EUV/X-ray stellar radiation,][]{Lammer03}.  The process of atmospheric erosion due to the impact of an ionized wind is complex, and here we simply examine the degree to which the planetary magnetosphere is penetrated by the wind. 

On this basis, a planetary magnetic field similar to that of the Earth seems to be strong enough to largely resist the stellar wind in the super-Alfv\'enic sectors. However, in the case of the sub-Alfv\'enic sectors there is no bow shock at all, and many field lines are connected directly from the stellar wind to the planet, resulting in poor atmospheric protection.  In a manner of fact, Planet A is located most of its orbit in a sub-Alfv\'enic plasma and the assumption of the existence of a bow shock \citep[e.g.,][]{griessmeier04,Khodachenko07,Lammer07,Vidotto11} may not be relevant for planets of this kind.

\cite{griessmeier04} have studied the atmospheric protection for hot-jupiter planets using scaling laws for the stellar mass loss and the planetary magnetic moment, and by estimating the atmospheric loss of neutral hydrogen. They concluded that hot-jupiters may have been significantly eroded during the early stages of the stellar system when stellar magnetic activity was high, and assuming that the mass-loss rate of young stars is comparatively high \citep[about 100 times the current solar value,][]{Wood02}.  Other estimates of the mass loss rates of young stars suggest that the winds might not be so strong \citep[only about 10 time the current Sun,][]{Holzwarth11,Sterenborg11}. In the case of the planets studied here, we do not include any atmospheric outflow. However, our simulations demonstrate that in order to estimate such an outflow and the degree of planetary protection afforded by a magnetic field, {\it one should consider not only the magnitude of the stellar wind pressure, but also whether the surrounding space plasma is sub- or super-Alfv\'enic} and the orientation of the stellar magnetic field, as it may allow stellar wind particles to flow directly into the planetary atmosphere.

\subsection{CHANGE IN MAGNETOSPHERIC STRUCTURE ALONG THE ORBIT}
\label{sec:ChangeInMagnetosphere}

The time-dependent results show that the magnetospheres of the planets under study experience significant changes in their topology and nature on timescales of a few hours, due only to their transition through different plasma conditions along their orbit. While the magnetospheres have an Earth-like structure during times when the planet passes through super-Alfv\'enic sectors, they have an Io-like Alfv\'en wing shape during times when the planet passes through sub-Alfv\'enic sectors.  Planet A and Planet B reside in the sub-Alfv\'enic sectors for most of their orbits.  Surprisingly, the super-Alfv\'enic sectors are associated with slower wind speed. These sectors are super-Alfv\'enic due to the order of magnitude increase in the plasma density, which reduces the Alfv\'en speed in the wind. 

On the Earth, the Cross Polar Cap Potential (CPCP) is the difference between the maximum and minimum ionospheric electric potential. It is associated with the solar wind driver and it becomes saturated during major CME events \citep[e.g.,][]{Reiff81,Siscoe02}. In the case of the planets studied here, the strong driving stellar wind is reflected in a saturated CPCP, where the magnitude of the CPCP is $10^2-10^4$ kV---much greater than a potential of about $50$ kV for the ambient solar wind conditions and $1000$ kV for CME conditions. \cite{Ridley07} and \cite{KivelsonRidley08} have discussed the dependence of the CPCP on the solar wind conditions. They showed that for larger $v_A$ (and likely sub-Alfv\'enic flow), the solar/stellar wind conductance, $\Sigma_A=(\mu_0 v_A)^{-1}$ ($\mu_0$ is the permeability of free space), is smaller than the ionospheric conductance, $\Sigma_p$, and as a result, the reflected fraction of the stellar wind electric field becomes larger than the incident one and the CPCP becomes saturated. In the case of our hypothetical planets, the stellar wind is always extremely strong. Therefore, it seems likely the CPCP in these planets is always saturated, unless $\Sigma_p$ is reduced even further due to the space environment conditions so it is again smaller than $\Sigma_A$. This can prevent from the stellar wind electric field to reflect and the CPCP is again in a non-saturated state. As mentioned in Section~\ref{sec:GMIE}, we leave the detailed calculation of the ionospheric conductance in close-in exoplanets to future study. 

\subsection{JOULE HEATING OF THE UPPER ATMOSPHERE}
\label{sec:JOULEHEATING}

Our results show that there is a significant Joule Heating at the upper atmospheres of the planets as a result of interaction with the extreme stellar wind. Overall, the heating is 2-5 orders of magnitude higher than that of the Earth during quiet solar wind conditions. The quiet-time heating for close-in exoplanets is even higher than those obtained on Earth during a strong CME event. It is most likely that the heating is even greater during CME events on these close-in planets as the energy deposited in such events can be 3 orders of magnitude larger than a typical CME on Earth \citep{Cohen11b}. The time-dependent simulation of Planet A shows that additional heating is available to the sharp and quick changes in the magnetospheric topology as the planet passes between the sub- to the super-Alfv\'enic sectors. In a way, these changes can be viewed as if a CME hits the planet twice in an orbit, leading to sharp and fast changes in the conditions of the driving stellar wind. 

As expected, the heating decreases with the increase of the planetary field strength as the planetary magnetic pressure reduces the stellar wind forcing. Of course, this can be modified by the particular orientation of the planetary and stellar magnetic fields as further heating can be driven by magnetic reconnection, which accelerates the precipitating electrons to higher energies \citep{RussellKivelson95,Gombosi99}. Our simulations also show that the heating increases with the decrease of the Pedersen conductance in the ionosphere. This is important because the conductance captures the role of the upper atmosphere in the energy input from the stellar wind. In reality, the conductance is defined by the atmospheric parameters and conditions. By estimating these parameters using a more detailed modeling \citep[e.g.,][]{Ridley02, Ridley04}, one can obtain a good estimation of the conductance and the overall heating of the top-side atmosphere by the interaction with the stellar wind.

While the Joule Heating of the upper atmosphere is far smaller than the planetary core Joule Heating necessary to explain the inflation of hot jupiter planets \citep{Batygin10,Perna10,Huang12,Menou12,Rauscher13,Spiegel13,Rogers14}, it is still significant in terms of the energy balance of the atmosphere. For the ambient stellar wind conditions, the heating of $10-50\; W\;m^{-2}$ can reach 0.1-3\% of the total stellar irradiating input shown in Table~\ref{table:t1}. As stated above, it can be even greater for periods of CMEs, as we expect the CME rate in M-dwarf stars to be high \citep{gudel07}. The additional heating at the top of the atmosphere is important for modelling the atmospheres of habitable planets \citep[e.g.,][]{Kasting93,Spiegel08,Heng11, Tian14} as they affect the atmospheric temperature. Therefore, these models should take in to account the atmospheric Joule Heating in the context of their pressure boundary conditions. In addition, Joule Heating at the top of the upper atmosphere transfers the energy to the thermosphere below, driving changes in the temperature, density, and pressure in the form of acoustic and gravity waves. As the changes in the magnetosphere and Joule Heating repeat along the orbit, it would be interesting to study the time-scale of the change in the driving force (i.e., the stellar wind conditions along the orbit), to the propagation time-scale of the planetary perturbations. However, this study is beyond the scope of this paper.

%%%%%%%%%%%%%%%%%%%%%%%%%%%%%%%%%%%%%%%%%%%%%%%%%%%%%%%%%%%%%%%%%%%%%%%%%%%%%%
% Section - Conclusions
%%%%%%%%%%%%%%%%%%%%%%%%%%%%%%%%%%%%%%%%%%%%%%%%%%%%%%%%%%%%%%%%%%%%%%%%%%%%%%

\section{SUMMARY AND CONCLUSIONS}
\label{sec:Conclusions}

In this paper, we study the magnetospheric structure and the ionospheric Joule Heating of habitable planets orbiting M-dwarf stars using a set of magnetohydrodynamic (MHD) models. The stellar wind solution is obtained using an MHD model for the stellar corona, which is driven by the magnetic field observations and the stellar parameters of EV Lac  - a mid-age M-dwarf star.  We investigate how the Joule Heating affects the upper atmospheres of three hypothetical planets located at the orbits of the three KOIs around EV Lac. We use the stellar wind conditions extracted at particular locations along the planetary orbit to drive an MHD model for the planetary magnetosphere, which is coupled with a model for the planetary ionosphere. The solutions from these simulations provide the magnetospheric structure and the Joule Heating of the upper atmosphere as a result of the interaction with the stellar wind.

Our simulations reveal the following major results:
\begin{itemize}
\item The space environment around close-in exoplanets can be very extreme, with the stellar wind dynamic pressure, magnetic field, and temperature being 1-3 orders of magnitude stronger than that at 1~AU. The stellar wind conditions along the planetary orbit change from sub- to super-Alfv\'enic.
\item The magnetosphere structure changes dramatically as the planet passes between sectors of sub- to super-Alfv\'enic plasma. While a bow shock is formed in the super-Alfv\'enic sectors, the planets reside in a sub-Alfv\'enic plasma most of the orbit, where no bow shock is formed and the stellar wind is directed towards the planetary surface. In this case, the protection of the planetary atmosphere is poor. 
\item A significant amount of Joule Heating is provided at the top of the atmosphere as a result of the planetary interaction with the stellar wind. The heating is enhanced in the time-dependent calculation as a result of the additional current due to the temporal changes in the magnetic field. For the steady-state, the heating is about 0.1-3\% of the total incoming stellar irradiation, and it is enhanced by 50\% for the time-dependent case. 
\item The transitioning between the plasma sectors along the planetary orbit has quantitive similarities to an exoplanet interacting with a CME.
\item The significant Joule Heating obtained here should be considered in models for the atmospheres of HZ planets in terms of the top-side temperature, density, and boundary conditions for the atmospheric pressure.
\end{itemize}

In this work, we have studied the interaction of magnetized habitable planets with the stellar wind. However, it is not clear whether the planetary magnetic fields of these planets are strong or weak. Alternatively, planets can have a thick enough, Venus-like atmosphere that can sustain the extreme stellar wind. We leave the investigation of such an interaction for future study.

%%%%%%%%%%%%%%%%%%%%%%%%%%%%%%%%%%%%%%%%%%%%%%%%%%%%%%%%%%%%%%%%%%%%%%%%%%%%%%
% Acknowledgments
%%%%%%%%%%%%%%%%%%%%%%%%%%%%%%%%%%%%%%%%%%%%%%%%%%%%%%%%%%%%%%%%%%%%%%%%%%%%%%

\acknowledgments

We thank an unknown referee for her/his comments and suggestions. The work presented here was funded by the Smithsonian Institution Consortium for Unlocking the Mysteries of the Universe grant 'Lessons from Mars: Are Habitable Atmospheres on Planets around M Dwarfs Viable?', and by the Smithsonian Institute Competitive Grants Program for Science (CGPS) grant 'Can Exoplanets Around Red Dwarfs Maintain Habitable Atmospheres?'. Simulation results were obtained using the Space Weather Modeling Framework, developed by the Center for Space Environment Modeling, at the University of Michigan with funding support from NASA ESS, NASA ESTO-CT, NSF KDI, and DoD MURI. The simulations were performed on the NASA HEC Pleiades system under award SMD-13-4076.  JJD was supported by NASA contract
NAS8--03060 to the {\em Chandra X-ray Center} during the course of this
research and thanks the Director, H.~Tananbaum, for continuing support
and encouragement. 

%%%%%%%%%%%%%%%%%%%%%%%%%%%%%%%%%%%%%%%%%%%%%%%%%%%%%%%%%%%%%%%%%%%%%%%%%%%%%%
% Bibliography
%%%%%%%%%%%%%%%%%%%%%%%%%%%%%%%%%%%%%%%%%%%%%%%%%%%%%%%%%%%%%%%%%%%%%%%%%%%%%%

\newpage

%\bibliographystyle{apj}
%\bibliography{KOIs.bib}

%%%%%%%%%%%%%%%%%%%%%%%%%%%%%%%%%%%%%%%%%%%%%%%%%%%%%%%%%%%%%%%%%%%%%%%%%%%%%%%
% Tables    
%%%%%%%%%%%%%%%%%%%%%%%%%%%%%%%%%%%%%%%%%%%%%%%%%%%%%%%%%%%%%%%%%%%%%%%%%%%%%%%

\begin{deluxetable}{ccccccc}
\tabletypesize{\footnotesize}
\tablecolumns{7}
\tablewidth{0pt}
\tablenum{1}

\tablecaption{Stellar and Planetary Parameters of the KOI Systems \label{table:t1}}
\tablehead{
\colhead{Planet} & \colhead{$R_\star \; [R_\odot]$} & \colhead{Stellar T$_{eff}\;[K]$} & \colhead{Semi-major Axis $[R_\star]$} & \colhead{$R_p\;[R_\oplus]$}  & \colhead{$Assumed\;B_p\;[G]$} & \colhead{$F\;[W\;m^{-2}]$}}
\startdata
A & 0.35 & 3482 & 36 & 1.37 & 0.3 & 1500 \\
B  & 0.22 & 3424 & 51.98 & 0.92 & 0.3 & 820\\
C & 0.4 & 3562 & 90 & 1.69 & 0.3 & 255
\enddata
\end{deluxetable}

\begin{deluxetable}{cccccccc}
\tabletypesize{\scriptsize}
\tablecolumns{8}
\tablewidth{0pt}
\tablenum{2}
\tablecaption{Stellar Wind Parameters used to drive GM \label{table:t2}}
\tablehead{
\colhead{Parameter}  & \colhead{Planet A} &  \colhead{Planet A} & \colhead{Planet B} &\colhead{Planet B} &\colhead{Planet C} &\colhead{Planet C} &\colhead{Planet C}\\
 &\colhead{sub-Alfv\'enic} &  \colhead{super-Alfv\'enic} & \colhead{sub-Alfv\'enic} &\colhead{super-Alfv\'enic} &\colhead{sub-Alfv\'enic} &\colhead{super-Alfv\'enic} &\colhead{super-Alfv\'enic}\\
& &   &  & &&\colhead{slow} &\colhead{fast} }
\startdata
$n\;[cm^{-3}]$ & 1100 & 34250 & 433 & 12895 & 46 & 3200 & 123  \\
$T\;[10^5\;K]$ &  $5.13$ & $8.37$ &$3.42$ & $4.77$ & $4.98$ &$2.22$ &$1.9$\\
$\mathbf{u}\;[km\;s^{-1}]$ &  (-609,-14,39) & (-140,101,13) & (-630,-1,30) & (-202,102,22) & (-728,-50,-17) & (-278,92,26) & (-660,8,14)\\
$\mathbf{B}\;[nT]$ &  (-1950,-377,170) & (-171,438,167) & (-804,-173,63) & (-57,223,92) & (240,88,17) & (-14,95,42) & (-244,74,18)\\
5$M_A$ &  0.46 & 2.95 & 0.73 & 4.76 & 0.88 & 7.25 &1.3
\enddata
\end{deluxetable}

\begin{deluxetable}{ccc}
\tabletypesize{\scriptsize}
\tablecolumns{8}
\tablewidth{0pt}
\tablenum{3}
\tablecaption{Solar Wind Parameters  \label{table:t3}}
\tablehead{
\colhead{Parameter}  & \colhead{Background Solar Wind} &  \colhead{CME Conditions} }
\startdata
$n\;[cm^{-3}]$ & 5 & 50 \\
$T\;[K]$ & $10^4$ & $5\cdot 10^4$ \\
$\mathbf{u}\;[km\;s^{-1}]$ & (-500,0,0) & (-1500,0,0) \\
$\mathbf{B}\;[nT]$ & (0,0,-5) & (0,0,-100) 
\enddata
\end{deluxetable}

%%%%%%%%%%%%%%%%%%%%%%%%%%%%%%%%%%%%%%%%%%%%%%%%%%%%%%%%%%%%%%%%%%%%%%%%%%%%%%%
% Figures    
%%%%%%%%%%%%%%%%%%%%%%%%%%%%%%%%%%%%%%%%%%%%%%%%%%%%%%%%%%%%%%%%%%%%%%%%%%%%%%%

%\begin{figure*}[h!]
%\centering
%\includegraphics[width=3.in]{EVLacBr}
%\caption{}
%\label{fig:f0}
%\end{figure*}

\end{document}